\documentclass[prd,a4paper,twocolumn,preprintnumbers,amsmath,amssymb,superscriptaddress,nofootinbib]{revtex4}

\usepackage{eufrak}

\begin{document}

\preprint{CERN-PH-TH-2009-254}
\preprint{CPHT-RR136.1209}

\title{Holomorphic variables in magnetized brane models with continuous Wilson lines}

\author{Pablo G. C\'amara}
\affiliation{CERN, PH-TH Division, CH-1211 Gen\`eve 23, Switzerland}

\author{Cezar Condeescu}
\affiliation{Centre de Physique Th\'eorique, Ecole Polytechnique and CNRS, F-91128 Palaiseau, France}

\author{Emilian Dudas}

\affiliation{Centre de Physique Th\'eorique, Ecole Polytechnique and CNRS, F-91128 Palaiseau, France}

\affiliation{LPT, Bat.210, Universit\'e de Paris-Sud, F-91405 Orsay, France}

\begin{abstract}
     We analyze the action of the target-space modular group in toroidal type IIB orientifold compactifications with magnetized D-branes and continuous Wilson lines. The transformation of matter fields agree with that of twisted fields in heterotic compactifications, constituting a check of type I/heterotic duality. We identify the holomorphic $\mathcal{N}=1$ variables for these compactifications. Matter fields and closed string moduli are both redefined by open string moduli. The redefinition of matter fields can be read directly from the perturbative Yukawa couplings, whereas closed string moduli redefinitions are obtained from D-brane instanton superpotential couplings. The resulting expressions reproduce and generalize, in the presence of internal magnetic fields, previous results in the literature.
\end{abstract}
%\pacs{xxxx}
\maketitle

%%%%%%%%%%%%%%%%%%%%%%%%%%%%%%%%%%%%%%%%%%%%%%%%%%%%%%%%%%%%%%%

\section{Introduction}

In the last decade, D-brane model building has become an outstanding approach for realizing semi-realistic chiral gauge theories within compactifications of String Theory \cite{review1,review2,review3}. Chirality in these constructions usually requires the presence of intersecting/magnetized D-branes. It is therefore of particular relevance to obtain expressions for the 4d effective action in these models. Efforts along that direction probably start with the computation of  Yukawa couplings in \cite{yukawa1,yukawa2,yukawa3,yukawa4}, and extend until the recently developed D-brane instanton calculus which allows the computation of non-perturbative superpotential couplings in these models (see e.g. \cite{revinst} and references therein). Moreover, higher order couplings can be expressed in terms of 3-point couplings \cite{yukawa2,yukawa3,higher1,higher2}, similarly to what happens in heterotic orbifold models. Thus, for many purposes, it suffices to study Yukawa couplings.

Whereas by now the picture that we have of the low energy effective action is fairly complete, there are still some aspects concerning the definition of $\mathcal{N}=1$ chiral variables in the presence of open string moduli which we believe deserve some further study.

More precisely, in a generic type II compactification the moduli space of open-string deformations (positions and Wilson lines of the D-branes) is non-trivially fibered over the closed string moduli space \cite{fibera1,fibera2,fibera3,fibera4}. Open string scalars enter in the definition of closed string moduli and, as a consequence, both kinds of deformations cannot be studied independently.

This becomes particularly important when analyzing the dynamics of D-branes within a closed string background, as required for instance in cosmological models of D-brane inflation \cite{inflation1,inflation2,inflation3,inflation4}. Although the explicit expression for these redefinitions is known in few examples \cite{fibera1,fiber1,fiber2,fiber3,fiber4,fiber5}, a systematic analysis in the context of D-brane model building is still missing. It is the purpose of this letter to fill in this gap.

Redefinition of the closed string moduli by the neutral D-brane scalars is related to the fact that closed string axions shift under some of the diagonal $U(1)$ gauge symmetries of the D-branes. Similarly, we can expect other charged fields, such as matter fields localized at intersections between different stacks of D-branes, to be also redefined by the D-brane moduli. We will find in Section \ref{yukawa} that this is indeed the case.

The structure of this letter is as follows. In Section \ref{yukawa}, we analyze the transformation of Yukawa couplings in magnetized brane models under fractional linear transformations of the complex structure and shifts of the Wilson line moduli. The transformation rules that we find agree with the ones obtained for twisted fields in heterotic orbifold compactifications \cite{heterotic1,heterotic2,heterotic3,heterotic4,heterotic5}. In particular, this allows us to identify the $\mathcal{N}=1$ chiral variables associated to charged matter fields in presence of arbitrary Wilson lines, in terms of which the superpotential is an holomorphic function.

Next, we consider the transformation properties of non-perturbative superpotential couplings induced by Euclidean D-brane instantons.\footnote{See \cite{grimm} for some related results in the context of $\mathcal{N}=1$ type IIB orientifold compactifications with O3 and O7-planes.} In this regard, in Section \ref{nonpert} we show that, after integration over the instanton moduli space, field redefinitions associated to instanton charged zero modes translate into redefinitions for the closed string K\"ahler and axion-dilaton moduli, thus making the full picture consistent.

Finally, we conclude with some last comments in Section \ref{concl}.

\section{Modular transformation of matter fields}
\label{yukawa}

Our starting point is the Yukawa coupling computation of Ref.\cite{yukawa4} (see also \cite{extra0,extra1,extra2,extra3}), performed by dimensionally reducing 10d $\mathcal{N}=1$ super Yang-Mills theory to 4d. Particularly important for us is the Wilson line dependence of this coupling. This dependence was determined by identifying the compact manifold with the moduli space of Wilson line deformations in the open string wavefunctions. The resulting expressions match the CFT computation performed in the type IIA side \cite{yukawa1}, up to global phases that we discuss below.

Hence, we consider $N=n_a+n_b+n_c$ D9-branes, with $n_\alpha\in\mathbb{N}^+$, and magnetization $m_{\alpha}\in\mathbb{Z}$ along a single $T^2$ given by,
\begin{equation}
F_{z\bar z}=\frac{\pi i}{\textrm{Im }\tau}\begin{pmatrix}\frac{m_a}{n_a}\mathbb{I}_{n_a}& &\\
&\frac{m_b}{n_b}\mathbb{I}_{n_b}&\\
&&\frac{m_c}{n_c}\mathbb{I}_{n_c}\end{pmatrix}
\end{equation}
where $\tau$ is the complex structure modulus of the torus and $\mathbb{I}_{n_\alpha}$ are $n_\alpha\times n_\alpha$ identity matrices. For simplicity, we take g.c.d.$(I_{ab},I_{bc},I_{ca})=\textrm{g.c.d.}(m_\alpha,n_\alpha)=1$, for each stack $\alpha=a,b,c$ of D9-branes, so that magnetization breaks the initial $U(N)$ gauge symmetry to $U(1)\times U(1)\times U(1)$. Intersection numbers are defined as
\begin{equation}
I_{\alpha\beta}\equiv m_\alpha n_\beta-m_\beta n_\alpha
\end{equation}
with $\alpha\beta=\{ab,bc,ca\}$, and determine the multiplicities of charged chiral multiplets, $\Phi^i_{\alpha\beta}$, $i=0\ldots |I_{\alpha\beta}|-1$, transforming in bifundamental representations of $U(1)\times U(1)\times U(1)$. Generalization to higher dimensional tori or non-abelian gauge groups is straightforward \cite{yukawa4}.

The expression for the physical Yukawa coupling among these charged fields is \cite{yukawa1,yukawa4},\footnote{Note that this expression (and therefore eq.(\ref{var}) below) is not exact. In particular, $\alpha'$ corrections are expected.}
\begin{multline}
Y_{ijk}=g\left(\frac{2\textrm{Im }\tau}{\mathcal{A}^2}\right)^{1/4}\mathcal{N}_{\mathcal{I}_{ab}}\mathcal{N}_{\mathcal{I}_{bc}}\mathcal{N}_{\mathcal{I}_{ca}}\\ \times e^{if_{ab}(\xi_{ab},\bar\xi_{ab})+if_{bc}(\xi_{bc},\bar\xi_{bc})+if_{ca}(\xi_{ca},\bar\xi_{ca})}\\ \times e^{i\pi\left(\frac{\xi_{ab}\textrm{Im }\xi_{ab}}{\mathcal{I}_{ab}}+\frac{\xi_{bc}\textrm{Im }\xi_{bc}}{\mathcal{I}_{bc}}+\frac{\xi_{ca}\textrm{Im }\xi_{ca}}{\mathcal{I}_{ca}}\right)/\textrm{Im }\tau}\\
\times \vartheta\left[{\delta_{ijk} \atop 0}\right](\xi;\tau|I_{ab}I_{bc}I_{ca}|)
\label{yuk}
\end{multline}
where $g$ is the 10d gauge coupling constant, $\mathcal{A}$ is the area of the 2-torus and  $\mathcal{N}_{\mathcal{I}_{\alpha\beta}}$ encodes the dependence of the K\"ahler metric on the intersection numbers. The concrete form of $\mathcal{N}_{\mathcal{I}_{\alpha\beta}}$ is not relevant for our purposes. The interested reader, however, can find explicit expressions in \cite{metric1,metric2,metric3,metric4,extra2}.

In addition, we are using the notation,
\begin{align}
\mathcal{I}_{\alpha\beta}&\equiv \frac{I_{\alpha\beta}}{n_\alpha n_\beta}\label{pio}\\
\delta_{ijk}&\equiv\frac{i}{{I}_{ab}}+\frac{j}{I_{ca}}+\frac{k}{I_{bc}}\\
\xi_{\alpha\beta}&\equiv\xi_\alpha-\xi_\beta\\
\xi&\equiv I_{ab}n_c\xi_c+I_{bc}n_a\xi_a+I_{ca}n_b\xi_b\label{xi}
\end{align}
where $\ \xi_{\alpha,x}$, $\xi_{\alpha,y}\in [0,1/n_\alpha)$ are the real Wilson line moduli along the 2-torus and $\xi_\alpha=-\xi_{\alpha,y}+\tau \xi_{\alpha,x}$. With these conventions the gauge potential associated to Wilson lines is given by $A_\alpha=\frac{i\pi}{\textrm{Im }\tau}\textrm{Im}(\xi_\alpha d\bar z)$.

As already mentioned, the CFT and field theory computations of the physical Yukawa coupling, performed respectively in \cite{yukawa1} and \cite{yukawa4}, leave undetermined an overall pure phase which depends on the Wilson line moduli. In particular, the results of \cite{yukawa1} and \cite{yukawa4} differ in such global phases. We have parameterized these terms in eq.(\ref{yuk}) by means of the unknown real functions  $f_{\alpha\beta}(\xi_{\alpha\beta},\bar\xi_{\alpha\beta})$. The need of these phases will become more clear below, when analyzing the transformation properties of the Yukawa coupling under periodic shifts of the Wilson line moduli. 

In order to consider supersymmetric compactifications, we must generalize the above example to the case where D9-branes are magnetized along three 2-tori, $T^2\times T^2\times T^2$. This can be easily done by taking three copies of eq.(\ref{yuk}), one for each 2-torus \cite{yukawa1,yukawa4}. Moreover, we shall take sign$(I^{(r)}_{ab}I_{bc}^{(r)}I_{ca}^{(r)})>0$. The physical Yukawa coupling can be then expressed in terms of a holomorphic superpotential,
\begin{equation}
W=\hat\Phi_{ab}^{\vec i}\hat\Phi_{bc}^{\vec j}\hat\Phi_{ca}^{\vec k}\prod_{r=1}^3\vartheta\left[{\delta_{i_rj_rk_r}^{(r)} \atop 0}\right](\xi^r;\tau_r|{I}_{ab}^{(r)}{I}_{bc}^{(r)}{I}_{ca}^{(r)}|)
\label{super}
\end{equation}
where the index $r$ refers to the $r$-th 2-torus (with analogous definitions to (\ref{pio})-(\ref{xi})) and we have taken the following expression for the holomorphic $\mathcal{N}=1$ chiral variables associated to matter fields,\footnote{Here, and in what follows, we neglect factors of $\mathcal{A}$, $g$ or $\mathcal{N}_{\mathcal{I}_{ab}}$, as they only play a role for modular transformations of the K\"ahler moduli or of the complex-axion dilaton, which we do not consider in this note.}
\begin{equation}
\hat\Phi_{\alpha\beta}^{\vec j}=e^{if_{\alpha\beta}}\left(\frac{ W}{\overline W}\right)^{\frac14}\left(\prod_{r=1}^3(\textrm{Im }\tau_r)^{\frac14}e^{i\pi\frac{\xi^r_{\alpha\beta}\textrm{Im }\xi^r_{\alpha\beta}}{\mathcal{I}^{(r)}_{\alpha\beta}\textrm{Im }\tau_r}}\right)\Phi_{\alpha\beta}^{\vec j}\label{var}
\end{equation}
The factor $(W/\overline{W})^{1/4}$ has its origin in the particular form of the $\mathcal{N}=1$ supergravity lagrangian, as explained in \cite{heterotic2}. In order to simplify the notation, we have combined the unknown phase factors of each of the three copies of (\ref{yuk}) into a single phase factor, $f_{\alpha\beta}\equiv \sum_{r=1}^3f^r_{\alpha\beta}$.

We want to analyze the transformation properties of this superpotential under linear fractional transformations of the complex structure. We consider, thus, the transformation $\tau_k\to \tau_k$ and $\tau_q\to -1/\tau_q$, for $q\neq k$ and $k$ fixed. This corresponds to T-dualizing along the four directions transverse to the $k$-th factor of $T^2\times T^2\times T^2$. Under this transformation the initial compactification is in general mapped into a different toroidal type IIB orientifold compactification. In particular, charges of D9-brane/O9-plane in the initial setup are mapped to charges of D5-brane/O5-plane wrapping the $k$-th 2-torus in the dual compactification, and viceversa. Similarly, magnetization in the D9-branes transforms as,
\begin{align}
(m^k_\alpha,n^k_\alpha)\ &\to \ (-m^k_\alpha,-n^k_\alpha)\ , \nonumber\\ (m^q_\alpha,n^q_\alpha)\ &\to \ (-n^q_\alpha,m^q_\alpha)\ , \quad q\neq k\label{transmag}
\end{align}

Let us analyze how the holomorphic variables in eq.(\ref{var}) transform under this duality. First, note that eq.(\ref{yuk}) as it stands is invariant under the above transformations.\footnote{Actually, one may check that it is also invariant under T-duality along a single 2-torus. For simplicity, however, here we just consider T-dualizing along two 2-tori.} From eqs.(\ref{super}) and (\ref{var}), and making use of Poisson re-summation and the discrete Fourier transform of the $\vartheta$-function \cite{yukawa4}, one may check that matter fields transform holomorphically as
\begin{equation}
\hat\Phi_{\alpha\beta}^{\vec j} \ \rightarrow \ \left(\prod_{q\neq k}\frac{e^{-\frac{\pi i (\xi'_{\alpha\beta})^2}{\mathcal{I}_{\alpha\beta}\tau}}}{\sqrt{-i\tau}}\right)\hat{\Phi}_{\alpha\beta}'^{\vec j}
\label{tran1}
\end{equation}
where in the T-dual setup, chiral variables are given in a new basis,
\begin{align}
\xi_\alpha'&=\xi_{\alpha,x}+\tau \xi_{\alpha,y}\label{bas1}\\
\hat{\Phi}'^{\vec j}_{\alpha\beta}&=\sum_{p_r=0}^{|{I}^{(r)}_{\alpha\beta}|-1}\sum_{p_s=0}^{|{I}^{(s)}_{\alpha\beta}|-1}\left(\prod_{q=r,s}\frac{e^{2\pi i pj\frac{{I}_{\beta\gamma}{I}_{\gamma\alpha}}{{I}_{\alpha\beta}}}}{\sqrt{|I_{\alpha\beta}|}}\right)\left.\hat \Phi^{\vec p}_{\alpha\beta}\right|_{\xi_{\alpha\beta}\to\xi_{\alpha\beta}'}\label{bas2}
\end{align}
with $r\neq s\neq k$. In order to simplify the notation in these expressions we have omitted the subindex $q$ in all quantities appearing within the brackets. Notice that, whereas eq.(\ref{tran1}) is a holomorphic field-dependent transformation, eq.(\ref{bas2}) is just a linear change of basis with constant phases as coefficients, which brings the superpotential back to its original form.

Eqs.(\ref{tran1})-(\ref{bas2}) have precisely the same structure than the corresponding ones for twisted fields in heterotic orbifold compactifications \cite{heterotic1,heterotic2,heterotic3,heterotic4,heterotic5}. The place of the orbifold twist  in heterotic orbifold compactifications is now taken by the quantity ${I}_{\beta\gamma}{I}_{\gamma\alpha}/{I}_{\alpha\beta}$, whereas the orbifold lattice corresponds in the type II language to the lattice generated by the intersections $k=0\ldots |{I}_{\alpha\beta}|-1$. Indeed, the fact that matter fields in type I compactifications with magnetized branes transform as twisted fields in heterotic compactifications is not surprising, as they are related by type I/heterotic S-duality.

Similarly, we can also consider the transformation of matter fields under shifts of the complex structure, $\tau_r\to \tau_r+1$. This has to be conveniently accompanied by a shift of the Wilson line moduli \cite{heterotic5},
\begin{equation}
\tau_r\to\tau_r+1\ , \ \ \ \textrm{Re }\xi^r_\alpha \to \textrm{Re }\xi^r_\alpha - \frac{\textrm{Im } \xi^r_{\alpha}}{\textrm{Im }\tau_r}\ ,\ \ \forall \alpha
\end{equation}
From eq.(\ref{var}), we observe that matter fields are invariant under this transformation. In order the superpotential to be also invariant, however, we have to define a new basis for the chiral variables, similarly to what we have just seen for the $\tau\to-1/\tau$ transformation. For ${I}^{(r)}_{ab}{I}^{(r)}_{bc}{I}^{(r)}_{ca}$ even, the new basis for matter fields is given by
\begin{equation}
\hat\Phi'^{\vec j}_{\alpha\beta}= e^{-i\pi j^2\frac{{I}_{\beta\gamma}{I}_{\gamma\alpha}}{{I}_{\alpha\beta}}}\hat\Phi^{\vec j}_{\alpha\beta}\label{shifttau}
\end{equation}
where, again, subindices $r$ should be understood in all quantities. This transformation is in agreement with the result for twisted fields in heterotic orbifold compactifications.\footnote{For ${I}^{(r)}_{ab}{I}^{(r)}_{bc}{I}^{(r)}_{ca}$ odd, however, the basis is more involved. In particular, in addition to eq.(\ref{shifttau}), one may check that $\vartheta\left[{\delta_{ijk} \atop 0}\right]$ is mapped to $\vartheta\left[{\delta_{ijk} \atop 1/2}\right]$ in eq.(\ref{super}). We have not succeeded in explaining this transformation in terms of a new basis for the matter fields. Similar problems under $\tau\to\tau+1$ were encountered in
heterotic orbifold compactifications when the order of the twist was even (see e.g. \cite{heterotic4}).}

Finally, let us consider the transformation of matter fields under shifts of the Wilson line moduli. Due to the periodicity of real Wilson lines, $\xi^r_{\alpha,x},\xi^r_{\alpha,y}\in [0,1/n^r_\alpha)$, the 4d effective action should be invariant under shifts $\xi^r_{\alpha,x(y)}\to\xi^r_{\alpha,x(y)}+1/n^r_\alpha$. Indeed, note that both $U(1)$ gauge transformations in the 4d theory and shifts of the Wilson line moduli have a common origin in $U(1)$ gauge transformations of the 10d gauge theory in the worldvolume of D9-branes. Gauge invariance of the 10d action then guarantees invariance of the 4d theory under shifts of the Wilson line moduli (up to possible change of basis within the space of degenerate fields).

In general, a shift in a Wilson line modulus induces a gauge transformation for the 4d fields which are charged under the corresponding $U(1)$. More precisely, after dimensional reduction, 4d fields are associated with eigenfunctions $\phi(x^r)$ of the Dirac operator on the internal torus. Torus translations induce $U(1)$ gauge transformations, $\phi(x^r+1)\to \textrm{exp}(iq\chi_r)\phi(x^r)$, with $q$ the $U(1)$ charge of the field and $\chi_r$ the corresponding Wilson loop. In the presence of Wilson lines, the eigenfunctions of the Dirac operator are actually $\phi(x^r+q\xi_{x^r})$, where $\xi_{x^r}$ denotes the Wilson line along the direction $x^r$ \cite{yukawa4}. Hence, a shift of the Wilson line modulus induces the same $U(1)$ gauge transformation than the corresponding torus translation.

This symmetry can be also understood in more geometrical terms in the mirror setup with intersecting D6-branes. In this picture, one of the real Wilson line moduli becomes the transversal distance of a D6-brane to the origin \cite{yukawa1}. The transformation $\xi^r_{\alpha,x}\to\xi^r_{\alpha,x}+1/n^r_\alpha$ then corresponds to move the D6-brane transversally along the $r$-th 2-torus until it reaches again the initial locus. Notice, however, that after moving the brane along this path intersections with other branes are permuted. In order to make explicit the invariance of the action, one has to apply a change of basis (in this case just a permutation) for the degenerate charged fields localized at different intersections. To avoid this sort of subtleties, here instead we choose to shift the modulus an enough number of times such that the permutation is trivial. This is achieved by,
\begin{equation}
\xi^r_{\alpha}\to \xi^r_{\alpha}+\delta^{r}_\alpha \label{xi1}
\end{equation}
or,
\begin{equation}
\xi_{\alpha}^r\to \xi_{\alpha}^r+\delta^{r}_\alpha\tau\label{xi2}
\end{equation}
with $\delta_\alpha^{r} =\textrm{l.c.m.}({I}^{(r)}_{\alpha\beta},{I}^{(r)}_{\alpha\gamma},\ldots)/n^r_\alpha$, where the numerator represents the lowest common multiple of the intersection numbers (along the $r$-th 2-torus) of brane $\alpha$ with all other branes present in the compactification.

We can easily check that physical Yukawa couplings are invariant under transformations (\ref{xi1}) and (\ref{xi2}) provided that matter fields transform, respectively, as
\begin{align}
\hat\Phi^{\vec i}_{\alpha\beta}\ &\to \ \hat\Phi^{\vec i}_{\alpha\beta}\ , \\
\hat\Phi^{\vec j}_{\gamma\alpha}\ &\to \ \hat\Phi^{\vec j}_{\gamma\alpha}\ ,
\end{align}
and,
\begin{align}
\hat\Phi^{\vec i}_{\alpha\beta}\ &\to \ e^{\frac{i\pi\delta^2_\alpha\tau}{\mathcal{I}_{\alpha\beta}}+\frac{2\pi i\xi_{\alpha\beta}\delta_\alpha}{\mathcal{I}_{\alpha\beta}}}\hat\Phi^{\vec i}_{\alpha\beta}\\
\hat\Phi^{\vec j}_{\gamma\alpha}\ &\to \ e^{\frac{i\pi\delta^2_\alpha\tau}{\mathcal{I}_{\gamma\alpha}}-\frac{2\pi i\xi_{\gamma\alpha}\delta_\alpha}{\mathcal{I}_{\gamma\alpha}}}\hat\Phi^{\vec j}_{\gamma\alpha} \ .
\end{align}
where subindices $r$ have been omitted. With the above definition (\ref{var}), this can be only the case if the phases $f_{\alpha\beta}$, $f_{\gamma\alpha}$, are non-trivial and transform under (\ref{xi1}) and (\ref{xi2}) as,
\begin{align}
f_{\alpha\beta}\ &\to \ f_{\alpha\beta}-\frac{\pi\delta_\alpha\textrm{Im }\xi_{\alpha\beta}}{\mathcal{I}_{\alpha\beta}\textrm{Im }\tau}\label{f1}\\
f_{\gamma\alpha}\ &\to \ f_{\gamma\alpha}+\frac{\pi\delta_\alpha\textrm{Im }\xi_{\gamma\alpha}}{\mathcal{I}_{\gamma\alpha}\textrm{Im }\tau}
\end{align}
and,
\begin{align}
f_{\alpha\beta}\ &\to \ f_{\alpha\beta}-\frac{\pi\delta_\alpha\textrm{Im }(\bar\tau\xi_{\alpha\beta})}{\mathcal{I}_{\alpha\beta}\textrm{Im }\tau}\\
f_{\gamma\alpha}\ &\to \ f_{\gamma\alpha}+\frac{\pi\delta_\alpha\textrm{Im }(\bar\tau\xi_{\gamma\alpha})}{\mathcal{I}_{\gamma\alpha}\textrm{Im }\tau} \ . \label{f2}
\end{align}
Consistently with this, one may verify that eq.(\ref{yuk}) is invariant under the transformations (\ref{xi1})-(\ref{xi2}) and (\ref{f1})-(\ref{f2}).

\section{Non-perturbative effects}
\label{nonpert}

\subsection{D-brane instantons}

We can extend without much effort the above reasoning to the case of non-perturbative superpotential couplings generated by D-brane instantons. The latter consist of Euclidean D1- and D5-branes (also dubbed E$1$- and E$5$-brane instantons) wrapping, respectively, complex compact curves and the full compact space.\footnote{However, in a given toroidal orbifold compactification in general only one of the two types of branes contributes to the superpotential.} The resulting superpotential couplings can be computed in terms of CFT scattering amplitudes by means of the recently developed D-brane instanton calculus \cite{inst1,inst2,inst3,inst4,inst5,inst6,inst7}.

In general, only $\mathcal{N}=1$ instantons with a spectrum of neutral zero modes  given by two fermions, $\theta_\rho$, $\rho=1,2$, and four scalars $x_\mu$, $\mu=0,\ldots,3$, parameterizing the $\mathcal{N}=1$ superspace, contribute to the 4d effective superpotential \cite{inst1,inst2,inst3,inst4}. This requires, among other things, that instantons wrap rigid cycles, and therefore couple to blow-up modes.\footnote{Instantons wrapping non-rigid cycles or preserving $\mathcal{N}\geq 2$ supersymmetry, however, may contribute to higher F-terms \cite{bw1,bw2,bw3,bw4,bw5,bw6}.}

The structure of instanton generated couplings factorizes into a product of disc amplitudes times the contribution of one-loop fluctuations around the instanton background, given by an exponential of annuli and M\"obius amplitudes. In this subsection we consider stringy instantons with no field theory analogous, i.e. instantons with no charged bosonic zero modes localized at the intersections with the D-branes present in the compactification. In that case the one instanton contribution reads \cite{inst1,inst6,inst7},
\begin{multline}
\langle\hat\Phi_{\alpha\beta}^1[\vec{x}_1]\hat\Phi_{\alpha\beta}^2[\vec{x}_2]\ldots\rangle=\\
\int d^4xd^2\theta\
 \sum_{\textrm{conf.}}\prod_{j=0}^{|I_{\alpha Ep}|-1} d\lambda^j_{\alpha} \prod_{k=0}^{|I_{Ep\beta}|-1}d\bar{\lambda}^k_{\beta}\ \cdot \mu(\xi_\alpha,\xi_\beta)  \\  e^{2\pi iS_{\textrm{E}p}} \langle\hat\Phi_{\alpha\beta}^1[\vec{x}_1]\rangle_{\lambda_\alpha,\bar\lambda_\beta}\cdot \langle\hat\Phi_{\alpha\beta}^2[\vec{x}_2]\rangle_{\lambda_\alpha,\bar\lambda_\beta} \cdot \ldots \times \\
\times e^{\mathcal{M}_{Ep}+\sum_s \mathcal{A}_{EpDq_s}}
\label{nonp}
\end{multline}
where the sum over configurations extends over all possible ways of distributing the charged fermionic zero modes, $\lambda_\alpha^j$ and $\bar{\lambda}^k_\beta$, among the disc correlators, $\langle\hat\Phi_{\alpha \beta}^r[\vec{x}_i]\rangle_{\lambda_\alpha,\bar\lambda_\beta}$.

Each correlator consists of a product of matter fields,
\begin{equation}
\hat\Phi_{\alpha\beta}^r[\vec{x}_i]=\hat\Phi_{\alpha x_{i}^1}\hat\Phi_{x_{i}^1x_i^2}\ldots \hat\Phi_{x_{i}^{n_r-1}x_i^{n_r}}\hat\Phi_{x_{i}^{n_r}\beta}
\end{equation}
and two charged zero modes, $\lambda_\alpha$ and $\bar\lambda_\beta$, laying at the $\textrm{E}p-\alpha$ and $\beta-\textrm{E}p$ sectors, respectively. Notice that, due to the factorization property of higher order couplings, any disc amplitude $\langle\hat\Phi_{\alpha \beta}^r[\vec{x}_i]\rangle_{\lambda_\alpha,\bar\lambda_\beta}$ can always be expressed in terms of 3-point disc amplitudes \cite{higher2}. We have also introduced a measure $\mu(\xi_\alpha,\xi_\beta)$ over the (charged) moduli space of the instanton. An explicit expression for this measure will be proposed in \cite{toappear} based on the invariance of eq.(\ref{factorE5}) under $U(1)$ gauge transformations and shifts of the Wilson line moduli. This expression, however, is not relevant for our purposes.

The last term in eq.(\ref{nonp}) contains contributions due to one loop fluctuations around the instanton background, where the sum runs over all D5- and D9-branes in the compactification. This term depends on the position and Wilson line moduli of branes which are parallel to the instanton along some direction (Neumann-Neumann or Dirichlet-Dirichlet boundary conditions) and which, therefore, develop extra massless fermionic modes at particular loci of their moduli space, corresponding to the zeroes of exp$({\mathcal{M}_{Ep}+\sum_s \mathcal{A}_{EpDq_s}})$ \cite{ganor}.

Finally, $S_{\textrm{E}p}$ is the tree-level part of the instanton action. For E5 or E1 instantons it is given respectively by,
\begin{equation}
S_{E5} = S + M_0 \ , \qquad S_{E1_k} = T_k + M_k \ , \quad k=1,2,3
\end{equation}
where $T_k$ is the K\"ahler modulus of the $T^2$ wrapped by the E$1_k$ instanton, $S$ the complex axion-dilaton, and $M_A$ are linear combinations of complex blow-up moduli with coefficients depending on the discrete Wilson lines, position and Chan-Paton charge of the instanton. For $T^2\times T^2\times T^2$ orientifolds with O9-planes \cite{fibera1,luis,gl},
\begin{equation}
S = C_6+ig_s^{1/2}\prod_{r=1}^3\textrm{Vol}_{r}\ , \quad T_k = C_{2,k}+ig_s^{-1/2}\textrm{Vol}_{k}
\end{equation}
with $\textrm{Vol}_{r}$ the volume of the $r$-th 2-torus ($r=1, 2, 3$), $g_s$ the string coupling constant and $C_6$ ($C_{2,k}$) the component of the RR 6-form (2-form) along the 6d compact space (the $k$-th 2-torus).

Let us proceed now to analyze how the holomorphic variables defined in eq.(\ref{var}) apply in this non-perturbative context. We are interested in closed string moduli redefinitons by the Wilson line moduli. To be more precise, consider the contribution of a single non-magnetized E5-instanton to a superpotential coupling which involves only fields transforming in symmetric or antisymmmetric representations of the gauge group, that is, fields localized at the intersection between branes $\alpha$ and their images $\alpha^*$ under the orientfold projection. We could have considered more general couplings, however, this simple case is enough to read the holomorphic variables for the closed string moduli. Indeed, taking $\beta=\alpha^*$ and
\begin{equation}
(m_{\alpha^*}^r,n_{\alpha^*}^r)=(-m_{\alpha}^r,n_{\alpha}^r)\ , \qquad \xi^r_{\alpha^*}=-\xi^r_{\alpha}\ , \quad r=1,2,3
\end{equation}
we see, from eq.(\ref{yuk}) (or rather, its generalization to the case of a $T^2\times T^2\times T^2$) and factorization of the disc amplitudes $\langle\hat\Phi_{\alpha \beta}^r[\vec{x}_i]\rangle_{\lambda_\alpha,\bar\lambda_\beta}$ into 3-point amplitudes, that eq.(\ref{nonp}) contains a Grassman integral of the form,
\begin{widetext}
\begin{equation}
e^{2\pi i\left(S+M_0\right)} \prod_{\{\alpha\}}e^{2iI_{\alpha E5}f_{\alpha E5}+2\pi i\sum_{r=1}^3c_\alpha^r\frac{\xi^r_{\alpha E5}\textrm{Im }\xi^r_{\alpha E5}}{\textrm{Im }\tau_r}}\int\prod_{j,k=0}^{I_{\alpha E5}-1}  d\lambda^j_{\alpha}d\bar{\lambda}^k_{\alpha^*} \ \lambda^j_{\alpha}\bar{\lambda}^k_{\alpha^*} \ \mu(\xi_\alpha) \cdot \ldots
\cdot \prod_{\{\tilde\alpha\}}e^{2\pi i \sum_{r=1}^3c^r_{\tilde\alpha}\frac{\xi^r_{\tilde\alpha E5}\textrm{Im }\xi^r_{\tilde\alpha E5}}{\textrm{Im }\tau_r}}\label{factorE5}
\end{equation}
\end{widetext}
where, without lost of generality, we have taken $I_{\alpha E5}>0$. Here, the coefficients $c^r_\alpha$ correspond to the RR charges of D$5$-brane which are induced in the worldvolume of D9-brane $\alpha$  by the magnetization,
\begin{equation}
c^r_\alpha =n^r_\alpha m^j_\alpha m^k_\alpha\ , \qquad r\neq j \neq k = 1,2,3
\end{equation}
The dots denote a holomorphic modular form of weight $-1$ with respect to the $SL(2,\mathbb{Z})^3$ modular group of fractional linear transformations of the complex structure moduli $\tau_r$, $r=1,2,3$. Finally, the last term in eq.(\ref{factorE5}) corresponds to contributions of one-loop fluctuations, given by the last term of eq.(\ref{nonp}). The absolute value of these contributions is related to the Green function of the Laplacian on the torus \cite{green,fiber3,fiber5,green2},
\begin{equation}
\mathcal{G}(\xi_{ab},\bar\xi_{ab})=-\textrm{log }\left|e^{\frac{i\pi\xi_{ab}\textrm{Im }\xi_{ab}}{\textrm{Im }\tau}}\frac{\vartheta\left[{\frac12 \atop \frac12}\right](\xi_{ab};\tau)}{\eta(\tau)}\right|^2\label{gf}
\end{equation}
with $\eta(\tau)$ the Dedekind $\eta$-function. Hence, the product over $\tilde\alpha$ contains the contribution of non-magnetized D$5$-branes wrapping the $r$-th $T^2$, with Wilson line modulus $\xi^r_{D5_r}$, and D9-branes with magnetization only along a $T^2\times T^2$ submanifold.

Similarly to what occurs for perturbative superpotential couplings, the non-holomorphic exponential prefactors in eq.(\ref{factorE5}) are absorbed into redefinitions of the holomorphic $\mathcal{N}=1$ variables. In the case at hand, however, instanton charged zero modes are not physical. Euclidean brane instantons do not span the 4d space-time dimensions and therefore the lagrangian does not contain kinetic terms for the fields $\lambda_\alpha^j$, $\bar\lambda_\beta^k$, which do not propagate. The exponential prefactors cannot be therefore absorbed into redefinitions of these fields, but rather of the closed string moduli in $S_{\textrm{E}p}$.

Considering exponential prefactors which are independent of the discrete Wilson lines of the instanton, $\xi^r_{E5}$, and combining disc and one-loop contributions, we find the following definition for the holomorphic $\mathcal{N}=1$ chiral variable associated to the axion-dilaton,
\begin{equation}
\hat S = S\ +\ \sum_{\{\hat \alpha\}}\sum_{r=1}^3c_{\hat \alpha}^r\frac{\xi^r_{\hat \alpha}\textrm{Im }\xi^r_{\hat \alpha}}{\textrm{Im }\tau_r}\label{ss}
\end{equation}
where $\{\hat\alpha\}$ now denotes all branes in the model, magnetized and non-magnetized.

Notice that $\hat S$ does not transform holomorphically under shifts of the Wilson line moduli, eqs.(\ref{xi1})-(\ref{xi2}), unless the RR 6-form which appears in the definition of $S$ transforms also. Let us show that this is indeed the case. For that aim, consider the following piece of the ten dimensional supergravity action,
\begin{equation}
\int[\textrm{Tr}(F_2\wedge F_2\wedge F_2\wedge A)\wedge F_3+F_7\wedge F_3]\label{gscoupl}
\end{equation}
with $F_3$ and $F_7$ the RR 3-form and 7-form field strengths ($F_7=*F_3$), respectively. Integrating over the internal 6-torus, and keeping track of the Wilson line moduli, we obtain the following couplings in the 4d effective action,
\begin{multline}
\frac16\int dx^4 \ \epsilon^{\mu\nu\rho\sigma}F_{\nu\rho\sigma}\left[ \textrm{Re}(\partial_\mu S)+\right.\\
\left.+\sum_{\alpha}\left(m^1_\alpha m^2_\alpha m^3_\alpha A_\mu^\alpha +\sum_{r=1}^3c_\alpha^r\frac{\textrm{Im}[\bar \xi^r_\alpha\partial_\mu\xi^r_\alpha]}{\textrm{Im }\tau_r}\right)\right]\label{gs}
\end{multline}
where $\epsilon_{\mu\nu\rho\sigma}$ is the 4d antisymmetric tensor.

The first term in the second line of (\ref{gs}) is the one responsible for the standard Green-Schwarz mechanism (see e.g. \cite{gerardo}). In order to cancel the anomalous transformation of this term  under $U(1)$ gauge transformations of the 4d theory, a shift of $\textrm{Re }S$ proportional to $m^1_\alpha m^2_\alpha m^3_\alpha$ is required. In presence of continuous Wilson lines in the compact space we see that an extra term arises. This extra term is indeed the responsible for the transformation of the RR 6-form potential under shifts of the Wilson line moduli,
\begin{align}
\xi^r_\alpha&\to\xi^r_\alpha+\delta_r^\alpha \ : &
S  &\to  S  - c^r_\alpha\frac{\delta_\alpha^r\textrm{Im }\xi^r_\alpha}{\textrm{Im }\tau_r}\label{shift1}\\
\xi^r_\alpha&\to\xi^r_\alpha+\delta_r^\alpha \tau_r\ :&
S   &\to  S  - c^r_\alpha\frac{\delta_\alpha^r\textrm{Im }(\xi^r_\alpha\bar\tau_r)}{\textrm{Im }\tau_r}\label{shift2}
\end{align}
This also matches the discussion in previous section, where we argued that shifts of the Wilson line modulus have an analogous effect than $U(1)$ gauge transformations in the 4d theory, due to their common origin in $U(1)$ gauge transformations of the 10d theory. The analogy, however, has to be taken with some care. Whereas the ``charge'' of the closed string axion under a 4d $U(1)$ gauge transformation is proportional to the charge of $\textrm{D}3/\overline{\textrm{D}3}$-brane induced by the magnetization, the ``charge'' under shifts of the Wilson line moduli is proportional to the induced D$5$-brane charges. Thus, there are situations where the axion does not transform under 4d $U(1)$ gauge transformations but it does transform under shifts of the Wilson line modulus. We will see in next section a particular example of this type.

Taking into account (\ref{shift1}) and (\ref{shift2}), now it is straightforward to verify that $\hat S$ indeed transforms holomorphically under eq.(\ref{xi1}) and (\ref{xi2}), providing a good consistency check of eq.(\ref{ss}).

Regarding the exponential terms in eq.(\ref{factorE5}) which depend on both the discrete parameters of the instanton and the Wilson line moduli of the D9-brane, we will argue in \cite{toappear} that they must be canceled with the measure over the charged moduli space of the instanton, $\mu(\xi_\alpha)$, in order eq.(\ref{factorE5}) to be invariant under $U(1)$ gauge transformations and shifts of the Wilson line moduli. Note that terms depending on the instanton parameters cannot be in any case associated to perturbative redefinitions of fields.

Finally, let us briefly comment on the redefinition of K\"ahler moduli. The simplest way to obtain these redefinitions is to T-dualize eq.(\ref{ss}) along the directions transverse to the $k$-th factor of $T^2\times T^2\times T^2$. Under this transformation, $\tau_{j}\to -1/\tau_{j}$, for $j\neq k=1,2,3$, and magnetization numbers transform as in eqs.(\ref{transmag}), as we already saw in the previous section.\footnote{The transformation rules for the magnetization numbers are actually not unique.  Acting with a generator of $\mathbb{Z}_2\times \mathbb{Z}_2$, where each $\mathbb{Z}_2$ reverses the magnetization numbers of a $T^2\times T^2$, leads to another set of valid transformations. The resulting E1-instantons differ in their Chan-Paton charge. Consistently, all of these T-duality transformations lead to the same expression for the redefined K\"ahler moduli.} Moreover, $S\to T_k$.
 We reproduce here the resulting expression,
\begin{equation}
\hat T_k = T_{k}\ -\ \sum_{\{\hat\alpha\}}\left[c_{\hat\alpha}^0\frac{\xi^k_{\hat\alpha}\textrm{Im }\xi^k_{\hat\alpha}}{\textrm{Im }\tau_k}-\sum_{p\neq q\neq k}^3c_{\hat\alpha}^q\frac{\xi^p_{\hat\alpha}\textrm{Im }\xi^p_{\hat\alpha}}{\textrm{Im }\tau_p}\right]\label{tt}
\end{equation}
where,
\begin{equation}
c_\alpha^0\equiv n^1_\alpha n^2_\alpha n^3_\alpha
\end{equation}
Under shifts (\ref{shift1}) and (\ref{shift2}) of the Wilson line moduli, the complex scalars $T_{k}$ transform respectively as,
\begin{align}
\xi^r_\alpha&\to\xi^r_\alpha+\delta_r^\alpha \ : & T_{k}  & \to  T_{k} + c^{k,r}_\alpha\frac{\delta_\alpha^r\textrm{Im }\xi^r_\alpha}{\textrm{Im }\tau_r}\, \\
\xi^r_\alpha&\to\xi^r_\alpha+\delta_r^\alpha \tau_r\ :& T_{k}  & \to  T_{k} + c^{k,r}_\alpha\frac{\delta_\alpha^r\textrm{Im }(\xi^r_\alpha\bar\tau_r)}{\textrm{Im }\tau_r}\ ,
\end{align}
where $c^{k,r}_\alpha=c^{0}_\alpha$ for $r=k$, and $c^{k,r}_\alpha=-c^{p}_\alpha$
for $r\neq k\neq p$.

\subsection{Gaugino condensation}

We have seen how to extract from superpotential couplings generated by stringy instantons the holomorphic variables associated to closed string moduli in the presence of magnetized D-branes and continuous Wilson lines. Similar statements hold for non-perturbative superpotentials originating from field theory instantons or gaugino condensation (equivalently, fractional gauge instantons \cite{gaugino}) in the worldvolume of some D-brane.

In what follows, we illustrate the application of these techniques to the $T^4/\mathbb{Z}_2 \times T^2$ BSGP orbifold model \cite{t1,t2}, consisting of 16 non-magnetized D9-branes and 16 D5-branes. Strictly speaking, this model does not contain complex rigid cycles, and therefore non-perturbative superpotentials are not generated. However, in the same spirit than \cite{fiber3}, we assume that extra ingredients in the compactification lead to complex rigid cycles at the singularities of the $\mathbb{Z}_2$-orbifold and, in particular, to $O(1)$ instantons contributing to the superpotential \cite{revinst}. Alternatively, we could think of this model as one of the $\mathcal{N}=2$ subsectors of a $T^6/\mathbb{Z}_2\times \mathbb{Z}_2$ orbifold with discrete torsion \cite{torsion1,torsion2,torsion3,torsion4}.

Hence, we assume that E1-instantons stuck at the singularities of the $\mathbb{Z}_2$ orbifold and wrapping the transverse $T^2$ in $T^4/\mathbb{Z}_2 \times T^2$ induce non-perturbative superpotential terms. In particular, we consider D5-branes to be fractional, so that they are stuck also at the singularities. In that case E1-instantons at the singularities correspond to fractional gauge instantons (or gaugino condensates) for the gauge theory living in the worldvolume of the D5-branes.

According to the discussion in previous sections, in this case there is some dependence of the non-perturbative superpotential  on the Wilson line moduli of the D9-branes along the $T^2$. We can make use of the Green function for the Laplace operator in the $T^2$, eq.(\ref{gf}), to compute the explicit expression of this superpotential \cite{fiber3,fiber5},\footnote{This superpotential (and its holomorphic expression, eq.(\ref{superfin})) differs from the ones presented in \cite{fiber3,fiber5} in that it contains additional terms of the form exp$(2\pi i \xi_{E1}\textrm{Im }\xi_{E1}/\textrm{Im }\tau)$ which arise from eq.(\ref{gf}). One may check that without these terms the superpotential would not transform as a modular form of weight -1 under $\tau\to-1/\tau$. These terms will lead to relative phases between various instanton contributions in eq.(\ref{superfin}).}
\begin{multline}
W=\sum_{\xi_{E1}}e^{\frac{2\pi i}{\sigma(\xi_{E1})} (T +M(\xi_{E1}))}\ \eta(\tau)^{-\frac{8+\sigma(\xi_{E1})}{\sigma(\xi_{E1})}}\ \times\\
\prod_{a=1}^{16}\left(e^{\frac{i\pi(\xi_{a}-\xi_{E1})\textrm{Im}(\xi_{a}-\xi_{E1})}{\textrm{Im }\tau}}\vartheta\left[{\frac12 \atop \frac12}\right](\xi_{a}-\xi_{E1};\tau)\right.\\
\left.e^{\frac{i\pi(\xi_{a}+\xi_{E1})\textrm{Im}(\xi_{a}+\xi_{E1})}{\textrm{Im }\tau}}\vartheta\left[{\frac12 \atop \frac12}\right](\xi_{a}+\xi_{E1};\tau)\right)^{\frac{1}{4\sigma(\xi_{E1})}}\label{super0}
\end{multline}
The sum in this expression extends over the discrete lattice of the $\mathbb{Z}_2$, spanned by the discrete Wilson line parameter of the E1-instanton,
\begin{equation}
\xi_{E1}=-\epsilon_2+\tau\epsilon_1\ , \qquad \epsilon_i=0,1/2
\end{equation}
The particular linear combination of blow-up moduli to which the instanton couples, here represented by $M(\xi_{E1})$, depends on $\xi_{E1}$. The K\"ahler modulus and the complex structure modulus of the $T^2$ are given, respectively, by $T$ and $\tau$, whereas $\xi_a$ is the corresponding Wilson line of the $a$-th D9-brane along this torus. The fractional charge of the instanton corresponds to the dual Coxeter number of the condensing gauge group $\sigma(\xi_{E1})$, which is related to the $\beta$-function coefficient of the gauge theory in the worldvolume of the D5-branes at the singularity where the instanton sits.

Under shifts of $\xi_a$, the complex scalar $T$ transforms as,
\begin{align}
\xi_a&\to \xi_a+1\ , & T&\to T-\frac14 \frac{\textrm{Im }\xi_a}{\textrm{Im }\tau}\label{pir1}\\
\xi_a&\to \xi_a+\tau\ , & T&\to T-\frac14\frac{\textrm{Im}(\bar\tau\xi_a)}{\textrm{Im }\tau}\label{pir2}
\end{align}
Note that in this case the Green-Schwarz mechanism does not induce shifts of $T$ under $U(1)$ gauge transformations ($m^1m^2m^3=0$), however, D9-branes in this model carry some charge of fractional D5-brane, leading to (\ref{pir1})-(\ref{pir2}).\footnote{This is also consistent with the mixing between $S$ and $T$ in the gauge kinetic function of D9-branes for this model found in \cite{sdual2}.}

We can write the global superpotential eq.(\ref{super0}) in terms of holomorphic variables, following the discussion in previous sections. More precisely, making use of the periodicity formula of the theta function, we express (\ref{super0}) as,
\begin{multline}
W=\sum_{\epsilon_i=0,\frac12}\eta(\tau)^{-\frac{8+\sigma(\xi_{E1})}{\sigma(\xi_{E1})}}\ e^{\frac{2\pi i}{\sigma(\xi_{E1})} (\hat T +M)}\times \\
e^{-\frac{8\pi i}{\sigma(\xi_{E1})}(\epsilon_2+\epsilon_1\epsilon_2)}\prod_{a=1}^{16}\vartheta\left[{\frac12+\epsilon_1 \atop \frac12+\epsilon_2}\right]^{\frac{1}{2\sigma(\xi_{E1})}}(\xi_{a};\tau)\label{superfin}
\end{multline}
where we have redefined the K\"ahler modulus \cite{fibera1},
\begin{equation}
\hat T= T+\frac14\sum_{a=1}^{16}\frac{\xi_a\textrm{Im }\xi_a}{\textrm{Im }\tau}\label{ttfin}
\end{equation}
Notice that blow-up moduli are not redefined in eq.(\ref{superfin}). This is due to the symmetrization of (\ref{super0}) under the orientifold action. In addition, notice that without (\ref{pir1})-(\ref{pir2}), eq.(\ref{ttfin}) would transform non-holomorphically.

In these holomorphic variables the superpotential takes a form which resembles much the expression of a superstring partition function. This is not surprising, since worldsheet instantons are related by S-duality to E1-instantons.\footnote{Indeed, the S-duality map can be explicitly used in this model to compute E1-instanton corrections to the effective action \cite{sdual1,sdual2}.}

One may check that eq.(\ref{superfin}) satisfies all the required global consistency conditions. Indeed, given the K\"ahler potential,
\begin{equation}
K=-\textrm{log}\left[(\hat T-\bar{\hat{T}})(\tau-\bar\tau)-\frac14\sum_{a=1}^{16}(\xi_a-\bar\xi_a)^2\right]
\end{equation}
the resulting scalar potential is not only invariant under the transformations (\ref{pir1})-(\ref{pir2}), but also under $\tau\to -1/\tau$ and $\tau\to \tau+1$, providing a good consistency check of these global expressions.

\section{Discussion}
\label{concl}

We have analyzed the transformation properties of perturbative Yukawa couplings in models with magnetized D9-branes under the action of the target-space modular group. More precisely, we have considered linear fractional transformations of the complex structure moduli and shifts of the Wilson line moduli of magnetized D9-branes. This analysis permitted us to identify the right definitions for the holomorphic $\mathcal{N}=1$ variables associated to matter fields and, using techniques of D-brane instanton calculus, the holomorphic variables for closed string moduli. Our main results can be summarized by eqs.(\ref{var}), (\ref{ss}) and (\ref{tt}).

The precise knowledge of these variables is required in any global analysis of the D-brane dynamics within a toroidal compactification \cite{toappear}. This is especially relevant for D-brane inflationary models, where inflation is driven by one or several open string moduli. In particular, in \cite{toappear} we will make use of these results to analyze the non-perturbative dynamics of D-branes in $T^6/\mathbb{Z}_2\times \mathbb{Z}_2$ orbifold models with discrete torsion.

There are several interesting aspects on which the results of this letter could be extended. The redefinitions we give here are valid for toroidal (or toroidal orbifold) models. It should be possible, however, to generalize these expressions and express them in terms of quantities which are well defined in arbitrary Calabi-Yau compactifications. We may also think that further ingredients such as closed string fluxes could also affect the redefinition of closed string moduli. In this regard, the effect of closed string fluxes on the open string wavefunctions has been analyzed recently \cite{open}, reflecting that the wavefunctions of the lightest modes associated to matter fields (in the regime where supergravity is a valid description) do not feel the effect of the closed string background flux. Based on this, therefore, we do not expect the redefinitions discussed here to be altered by the presence of closed string fluxes in a globally consistent model.

\begin{acknowledgments}
We thank M. Berg, F. Marchesano, J. F. Morales, M. Schmidt-Sommerfeld and A. Uranga for useful comments and discussions. E.D. thanks Aspen
Center for Physics and P.G.C. is grateful to Ecole Polytechnique for hospitality during the completion of this work. Work supported in part by the European ERC Advanced Grant 226371 MassTeV, by the CNRS PICS no. 3059 and 4172, by the grants ANR-05-BLAN-0079-02, and the PITN contract PITN-GA-2009-237920.
\end{acknowledgments}

\end{document}